# Synthesis and Characterization of Nanocomposites of Fe Nanoparticles and Activated Carbon

Satyendra Prakash Pal[1,2*] and P. Sen[1]

[1]School of Physical Sciences, Jawaharlal Nehru University, New Delhi-110067, India
[2]Department of Physical sciences, Indian Institute of Science Education and Research Mohali, Knowledge city, Sector 81, SAS Nagar, Manauli-140306, Punjab, India
[*]sppal85@gmail.com

**Abstract**

We have synthesized the bulk form of the superparamagnetic nanocomposites of Fe nanoparticles and activated carbon. Here highly porous feature of activated carbon and magnetic properties of Fe nanoparticles were combined together to form the nanocomposites by mechanical mixing of different proportion of both of them. These nanocomposites are characterized by various characterization techniques like X-ray diffraction (XRD), transmission electron microscopy (TEM), magnetization measurement and Mössbauer spectroscopy. The observed XRD patterns have shown most intense peaks at $2\theta = 44.8^0$, which correspond to the (110) plane of the pure Fe in body centered cubic phase. Average particles sizes obtained by TEM imaging are in the range of 20 nm, which are below the required limit for Fe nanoparticles to be in superparamagnetic phase. The superparamagnetic nature of the nanocomposites together with trace amount of the ferromagnetic phase was confirmed by magnetization measurements and Mössbauer spectroscopy. Particles size calculated by the Langavin function fitting of the magnetization vs magnetic field curves are in qualitatively good agreement with the size obtained from TEM imaging.

# INTRODUCTION

Nanosized magnetic materials and their composites have unique electrical, chemical, structural, and magnetic properties, with potential applications in various fields [1-4]. The superparamagnetic nature of nanoparticles at room temperature has got suitable application in magnetic separation applications [5]. The development of superparamagnetic composites with large surface area and high porosity would be of significant interest and would extend the applications of magnetic separation techniques in biomedicine, catalysis and waste treatment [6]. The route to design of new materials consists of a clever mix of materials with unique properties. Activated carbon and iron are two such materials, the former a catalyst and the latter a magnetic solid. Carbon–iron based nanocomposite systems are of growing interest due to their improved magnetic properties with potential in sensor applications, catalysis and metallurgy. On the basis of its considerable demand, a prospective reduction of costs of these materials is envisaged if bulk quantities can be produced [7-8].

In this work we have synthesized the different nanocomposites of Fe nanoparticles, prepared as bare particles, and activated carbon. This unique bulk preparation methodology whereby the nanomaterial is directly fused to carbon employing special properties of the nanomaterial surface, addresses the utility of these materials in real applications. The highly porous feature of activated carbon and magnetic properties of Fe nanoparticles were combined together to form the nanocomposites by mechanical mixing of different proportion of both of them. The nanocomposites are characterized by XRD, TEM, magnetization measurement and Mössbauer spectroscopy.

**Experimental**

Fe nanoparticles have been synthesized by employing a novel, physical, top-down approach of electro explosion of wires (EEW) [9, 10]. Different nanocomposites of activated carbon and Fe nanoparticles were obtained by mechanical mixing of activated carbon and Fe nanoparticles, with different weight ratios by grinding together in a mortar and pestle. In this way three nanocomposites were obtained by grinding: (1) 33% of Fe nanoparticles and 66% of activated carbon, by weight, denoted as (1:2), (3) 28% of Fe nanoparticles and 72% of activated carbon, denoted as (1:2.5), and (4) 25% of Fe nanoparticles and 75% of activated carbon, denoted as (1:3). Each sample was monitored by employing X-ray diffraction (XRD), transmission electron microscope (TEM), magnetization measurements and $^{57}$Fe Mössbauer spectroscopy. XRD patterns were recorded on a PANalytical X'pert PRO diffractometer using Cu K$_\alpha$ radiation ($\lambda$ = 1.5418A$^0$). For TEM investigations a small drop of the diluted suspension was put on a carbon coated copper grid. After drying the grid, TEM characterization was carried out employing a JEOL 2100F machine. The magnetic properties were measured using a Quantum Design physical property measurement system (PPMS). The Mössbauer spectrum was recorded with a conventional constant acceleration spectrometer in transmission geometry using a $^{57}$Co/Rh source. The experimental setup was calibrated using a standard α–Fe foil at room temperature.

**Result and Discussions**

XRD analysis of nanocomposites shows a peak at 2θ = 44.8$^0$, which is due to Fe nanoparticles, to be most intense while the others are hardly visible. In the XRD spectrum reported in Fig.1, peaks position of Fe nanoparticles matches with those from bulk Fe in bcc phase and correspond to (hkl) planes (110). There are very weak XRD peaks at 2θ=65.0$^0$, 82.4$^0$, i.e., they are suppressed. Due to the nonequilibrium nature of the synthesis process the planes of the Fe nanoparticles gets reoriented [11].The peak at $2\theta=44.8^0$ in XRD spectra of each nanocomposite shows the presence of iron in each composites.

TEM images from different nanocomposites are presented in Fig. 2. TEM images give poor contrast as the weight percentage of activated carbon in the nanocomposites increases. Particle size histograms are shown in inset to the images. From the particle size histograms, calculated values of the particle sizes are 13.8nm, 19.83 nm, and 20.18nm for (1:2) composite, (1:2.5) composite, and (1:3) composite, respectively. Hence the particle size increases with increment of activated carbon weight percentage. It seems like the interconnected pores of the activated carbon provides the van der Waal interactions between the nanoparticles, to form the clusters. TEM images show that in all nanocomposites most of the particles have size ≤20 nm. Superparamagnetism is often observed for iron nanoparticles below about 20 nm size. Selected area electron diffraction patterns shown in insets confirm the crystal planes obtained by XRD spectra.

All the nanocomposites, generated by the composite preparation conditions described so far, can be attracted by a permanent ferrite magnet. On withdrawal of the magnetic field, the particles revert to their original arrangement. In order to confirm the magnetic state of the composites we performed different magnetic measurements, data are shown in Fig. 3 which consist of their hysteresis data (M-H curves), taken at 300K (Figs. 3b, d, f), ZFC and FC magnetization measurements (Figs. 3 a, c, e). For the ZFC experiment, the samples were cooled from room temperature to 2K, in the absence of a magnetic field. On doing so, the particle moments are blocked progressively along their anisotropy directions. After reaching T = 2K the magnetization is recorded during warm up in the presence of an external field of 200Oe. While the field cooled data has been acquired by cooling the samples in an external field of 200Oe. The observed temperature-dependent magnetization curves are plotted in Fig. 3. The temperature at which the ZFC curve exhibits a cusp is defined as the blocking temperature ($T_B$). From our data, a broad peak at~ 230 K corresponds to the blocking temperature of the nanocomposite system. The saturation magnetization ($M_S$), remanence magnetization (($M_R$), and coercivity values, obtained from the M-H curves, are given in Table. The $M_S$ value for the composites decreases as dilution with carbon is incresead and their coercivity values increases with increased dilution. The increase of coercivity in the composite may arise due to complex interactions, which can create strong pinning centres for the core moments during demagnetization [12]. So the M-H curves show the presence of small ferromagnetism together with superparamagnetism at room temperature.

Magnetic nanoparticle sizes can be calculated by using the Langevin equation, describing assemblies of particles with freely rotating moments [13]:

$$M/M_S = \coth(\alpha) - 1/\alpha = L(\alpha) \qquad \text{-------------------------------------------- (1)}$$

$L(\alpha)$ is called Langevin function.

where, $\alpha = \mu H/K_B T$

$M/M_S$ = magnetization (M) normalized to the saturation magnetization ($M_S$)

$k_B$ = Boltzmann's constant , $\mu$ = magnetic moment of the particle

Magnetic moment ($\mu$) and the diameter of a particle (d) are related as:

$$\mu = M_S d^3 \pi/6 \qquad \text{-------------------------------------------- (2)}$$

By using the above equations, we have calculated the size of the nanoparticles by Langevin function fitting of the M-H curves for (1:2), (1:2.5), and (1:3) nanocomposites, taken at 300K. Fig. 4 shows the fitted experimental data and Table sums the fitting results together with the experimental data. From Fig. 4 and Table, the fittings agree well with the experimental data. For different composites, the particle sizes obtained by Langevin function fitting, namely, 15.02, 18.80 and 18.32nm, are close to the ones estimated employing TEM and reported in Fig.2. The particle sizes calculated by magnetization data ($D_m$) are however smaller than the particle sizes observed from TEM measurement ($D_{TEM}$). The difference between $D_m$ and $D_{TEM}$ is most likely due to contributions of a magnetically "dead" layer reported to be present on the surface of the particles [14]. In this case, $D_m$ measures the magnetic particle; hence the dead layer has no magnetic component.

The room temperature Mössbauer spectrum of the (1:2.5) composite is presented next, shown in Fig. 5 and is fitted by a doublet and a sextet. Doublet shows the superparamagnetic part due to γ-Fe phase [15] and sextet, with value of isomer shift =0.000 mm/sec, quadrupole splitting =0.008 mm/sec and hyperfine field=33.0T, which are typical values for the presence of ferromagnetic α–Fe phase. The existence of the α–Fe phase is also clear from the XRD data of the composite, with a peak at $2\theta = 44.8^0$ which is the sole noticeable peak. This shows that the nanocomposite at room temperature are superparamagnetic in nature, but a small ferromagnetic component is still present, which is also seen in magnetic measurements, by the presence of a small hysteresis in the M-H curves [16].

**Conclusions**

We have been able to synthesize the bulk quantities of different nanocomposites of activated carbon and iron nanoparticles. XRD spectra confirms the presence of bcc phased Fe in each composite and TEM images show Fe naoparticles with size of ~20 nm, disperessed in carbon matrix. The composites are attracted by permanent ferrite magnet; magnetization measurements and Mössbauer spectra show the presence of trace amount of ferromagnetism together with superparamagnetism. These nanocomposites could be used for the various technological applications like magnetic separation.

**Acknowledgments**

We thank AIRF, JNU for XRD, TEM and Dr. V.P.S. Awana, NPL, Delhi for help with magnetization measurement. Mössbauer spectroscopy was performed at UGC-DAE CSR, Indore. S. P. PAL thanks CSIR, India for research fellowship.

**Figure legends:**

**FIGURE 1.** XRD lines from: (a) (1:2), (b) (1:2.5), and (c) (1:3) nanocomposites.

**FIGURE 2.** Transmission electron micrographs: (a) (1:2), (b) (1:2.5), and (1:3) nanocomposite. Upper insets show electron diffraction patterns, lower insets show histogram of particle size distribution.

**Figure 3.** ZFC and FC magnetization measurement curves for: (a) (1:2) nanocomposite, (c) (1:2.5) nanocomposite and (e) (1:3) nanocomposite respectively, while (b), (d) and (f) are their M-H curves, respectively, taken at 300K.

**Figure 4.** Hysteresis curves fitted by Langevin function: (a) (1:2), (b) (1:2.5), and (c) (1:3) nanocomposites.

**FIGURE 5.** $^{57}$Fe Mössbauer spectra of (1:2.5) composite: circles (experimental data), lines (fitted curve).

**Table legend:**

**Table.** M-H curve measurements data taken at 300K and Langevin function fitting parameters

**Figure 1:**

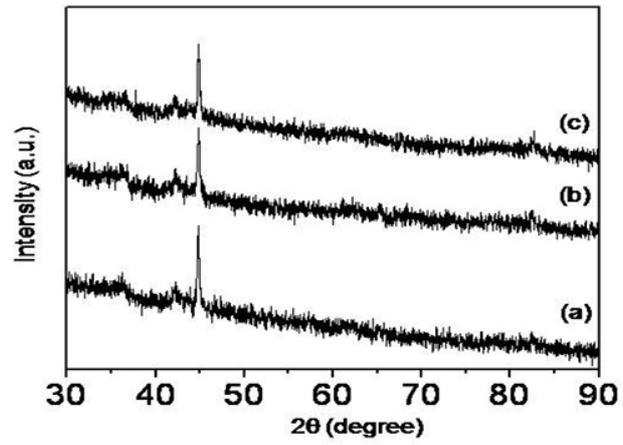

**FIGURE 2:**

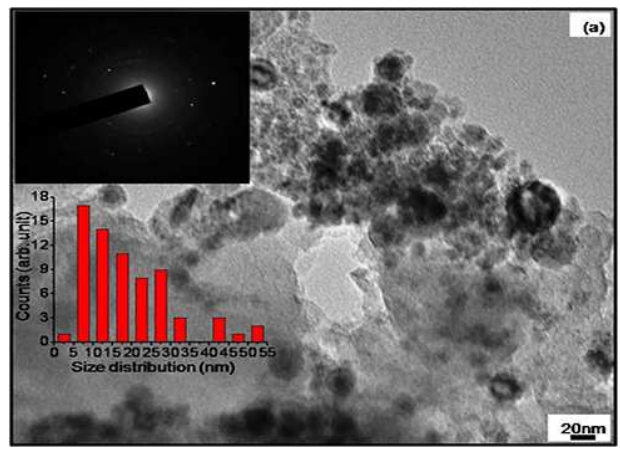

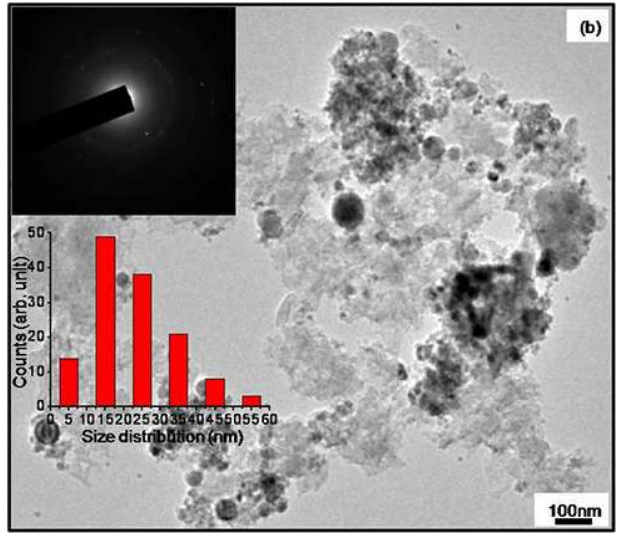

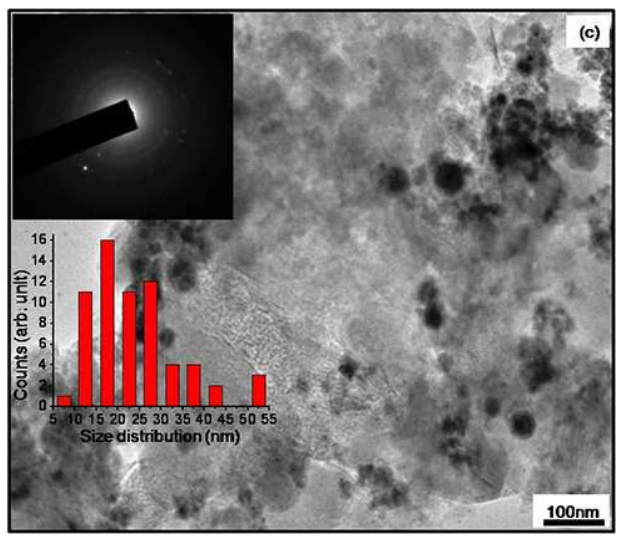

**FIGURE 3:**

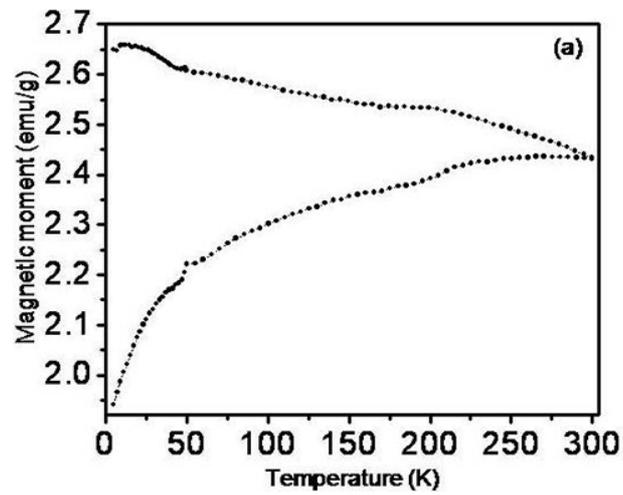

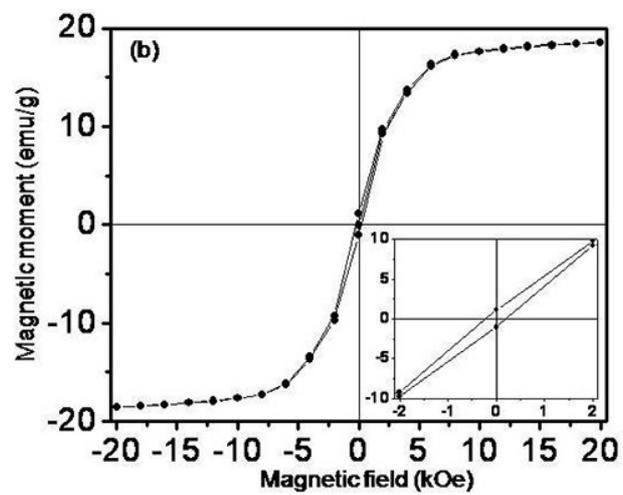

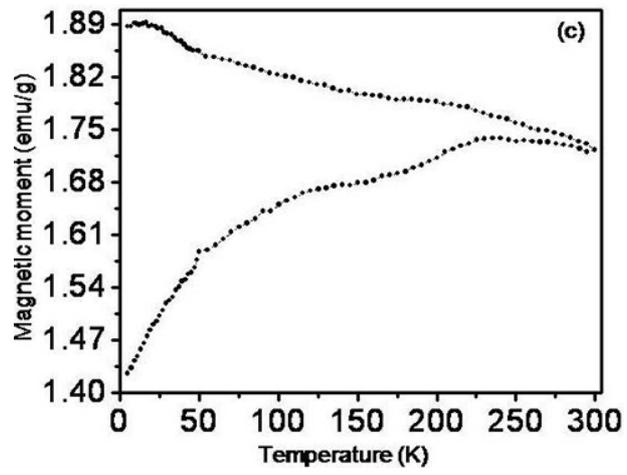

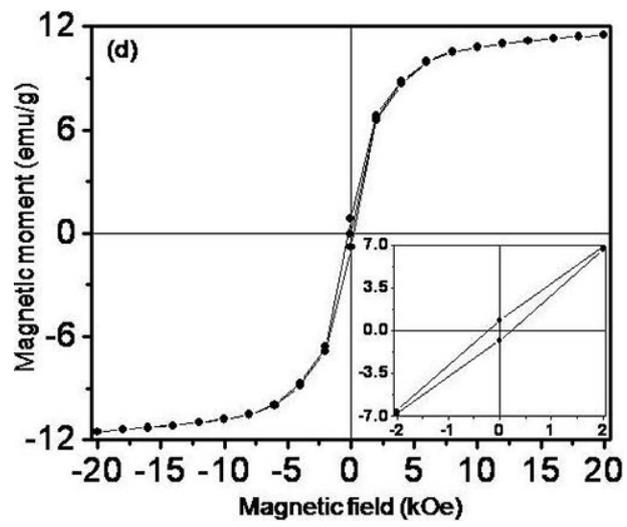

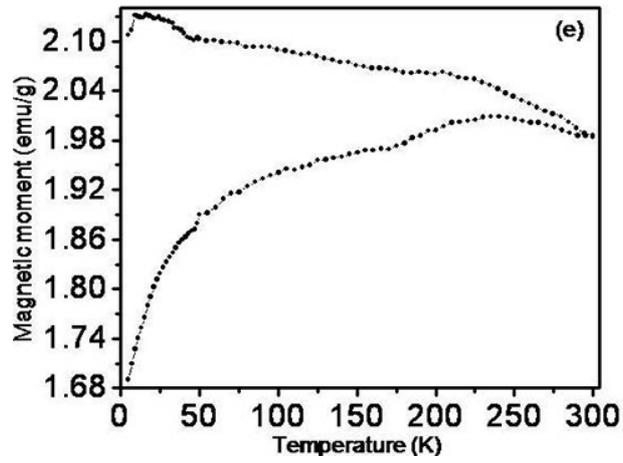

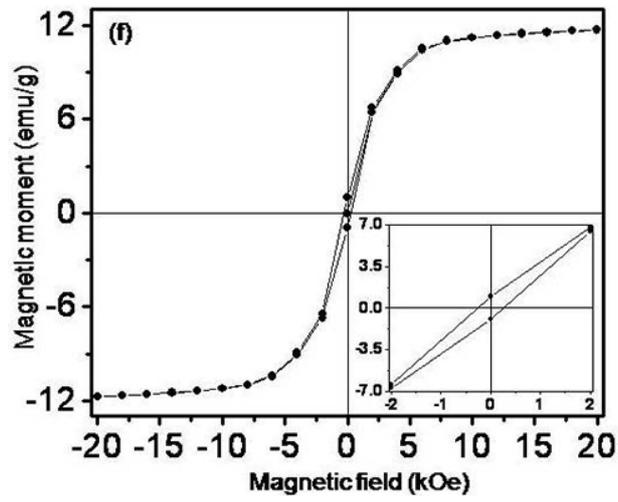

**FIGURE 4:**

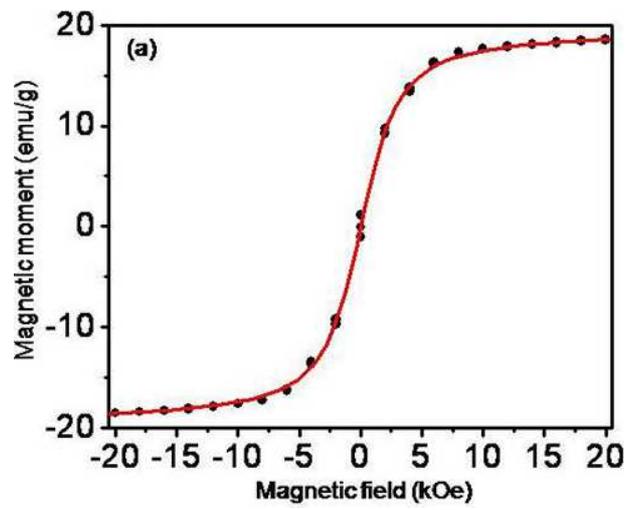

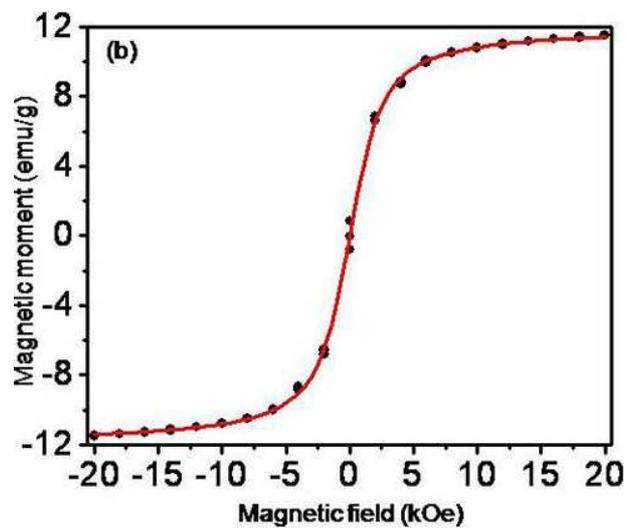

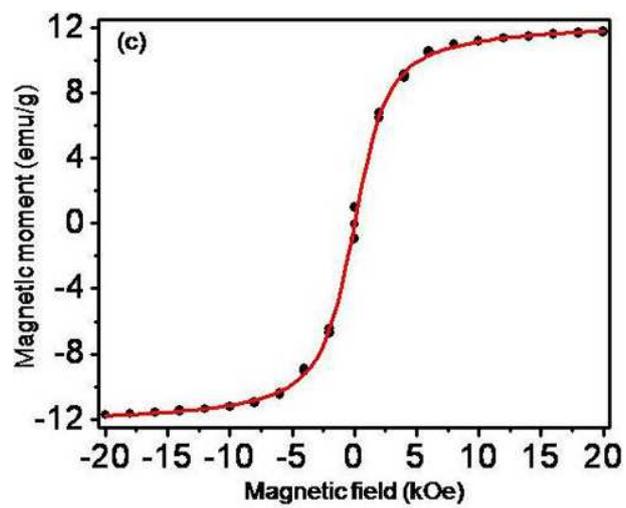

**FIGURE 5:**

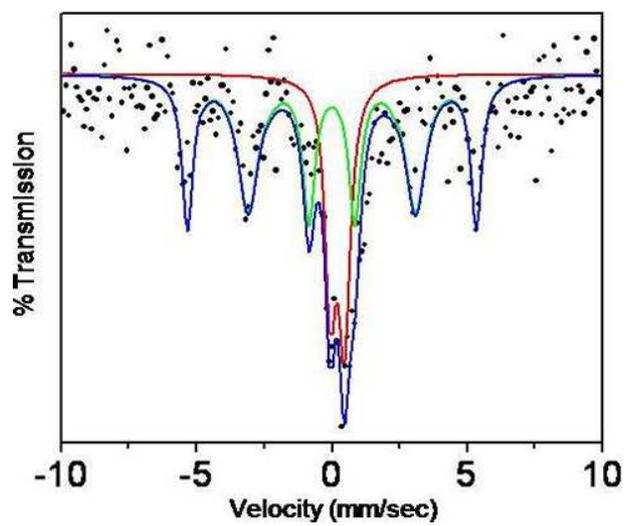

**Table:**

| Sample | $M_S$ (emu/g) | $M_R$ (emu/g) | $H_C$ (Oe) | Langevin Function Fitting | |
|---|---|---|---|---|---|
| | | | | d (nm) | $M_s$ (emu/g) |
| Composite (1:2) | 18.57 | 1.10 | 208.49 | 15.02 | 19.77 |
| Composite (1:2.5) | 11.42 | 0.81 | 220.41 | 18.80 | 12.01 |
| Composite (1:3) | 11.71 | 0.97 | 257.64 | 18.32 | 12.43 |